 \documentclass[aps,prl,twocolumn,groupedaddress]{revtex4}  
 \usepackage{epsfig}                                        
\usepackage{amsmath}
\usepackage{amsfonts}
\usepackage{amssymb}

\begin{document}

\title{
Spectroscopic studies of $^{168,170}$Dy using CLARA and PRISMA
}  

\author{
P.-A.~S\"{o}derstr\"{o}m$^1$,
J.~Nyberg$^{1}$,
P.~H.~Regan$^{2}$,
A.~Algora$^{3}$,
G.~de~Angelis$^{4}$,
S.~F.~Ashley$^{2}$,
S.~Aydin$^{5}$,
D.~Bazzacco$^{5}$,
R.~J.~Casperson$^{6}$,
W.~N.~Catford$^{2}$,
J.~Cederk\"{a}ll$^{7}$,
R.~Chapman$^{8}$,
L.~Corradi$^{4}$,
C.~Fahlander$^{9}$,
E.~Farnea$^{5}$,
S.~J.~Freeman$^{10}$,
A.~Gadea$^{3,4}$,
W.~Gelletly$^{2}$,
E.~Grodner$^{4}$,
C.~Y.~He$^{4}$,
G.~A.~Jones$^{2}$,
K.~Keyes$^{8}$,
M.~Labiche$^{8}$,
X.~Liang$^{8}$,
Z.~Liu$^{2}$,
S.~Lunardi$^{5}$,
N.~M\u{a}rginean$^{4,11}$,
R.~Menegazzo$^{5}$,
D.~Mengoni$^{5}$,
D.~Napoli$^{4}$,
J.~Ollier$^{8}$,
S.~Pietri$^{2}$,
Zs.~Podoly\`{a}k$^{2}$,
F.~Recchia$^{4}$,
E.~Sahin$^{4}$,
J.~F.~Smith$^{8}$,
K.-M.~Spohr$^{8}$,
S.~J.~Steer$^{2}$,
A.~Stefanini$^{4}$,
N.~J.~Thompson$^{2}$,
G.~Tveten$^{7,12}$,
C.~A.~Ur$^{5}$,
J.~J.~Valiente-D\'{o}bon$^{4}$,
V.~Werner$^{6}$,
S.~J.~Williams$^{2}$
\vspace{0.3cm} 
}

 \affiliation{                                              
1 Uppsala University,
2 University of Surrey,
3 IFIC-Valencia,
4 INFN-Legnaro,
5 INFN-Padova,
6 Yale University,
7 CERN,
8 The University of the West of Scotland,
9 Lund University,
10 University of Manchester,
11 IFIN HH-Bucharest,
12 University of Oslo
}

\maketitle 
\thispagestyle{empty} 

\pagestyle{empty} 

\section{Introduction}


Recent studies of neutron-rich exotic nuclei have been focused on the shell closure aspects and the behavior of the magic numbers on the neutron-rich side of the line of $\beta$ stability. Even more rare than the doubly magic nuclei, are the doubly mid-shell nuclei, with maximum number of valence protons and neutrons. The importance of the product of valence nucleons, $N_{\mathrm{p}}N_{\mathrm{n}}$, for quadrupole collectivity is well known, as both the energy of the first $2^{+}$ state and the energy ratio $4^{+}/2^{+}$ have a smooth dependence on this quantity \cite{casten0,casten1,mach,zhao}. This naively implies that the valence maximum, $^{170}$Dy, would be the most collective of all nuclei, but it has been discussed whether this is the case or not \cite{mach,166dy}. In fact the $\gamma$-ray energies of the ground state rotational bands in the neutron-rich dysprosium isotopes decrease up to $^{164}$Dy, but increase again for $^{166}$Dy. This has been interpreted as the maximum deformation actually occurring at $^{164}$Dy. However, the only spectroscopic measurement published on $^{168}$Dy to date, from a $\beta$-decay experiment, shows a decrease in the energies of the $2^{+}$ and $4^{+}$ states relative to $^{166}$Dy, suggesting that the energy increase in $^{166}$Dy is an irregular behavior \cite{168dy}. One question is does this decrease continue to higher spin in $^{168}$Dy and also for heavier even-even dysprosium isotopes. The neighbouring even-$Z$ elements in this region have a minimum of their $2^{+}$ state energy at the midshell $N=104$, corresponding to $^{170}$Dy. It would be natural to expect the same for the dysprosium isotopes.

Dysprosium is also well known for the original discovery of the backbending phenomenon \cite{backbending}. Total Routhian Surface calculations suggest a very deep, prolate, axially symmetric minimum for $^{170}$Dy \cite{paddy}. Increasing the spectroscopy to higher spin, up to and beyond the backbending, will give information regarding the stiffness of the  quadrupole deformation in these neutron-rich dysprosium isotopes.


\section{Experimental setup}

The experiment reported here was carried out using multi-nucleon transfer reactions between $^{82}$Se and $^{170}$Er. The beam was $^{82}$Se at an energy of 460 MeV and an intensity of $\sim$25~enA ($\sim$2~pnA), delivered by the Tandem XTU-ALPI accelerator complex at LNL. This beam was incident on a 500 $\mu$g/cm$^{2}$ thick self-supporting $^{170}$Er target. Beam-like fragments were identified using the PRISMA magnetic spectrometer \cite{prisma}, placed at the grazing angle of 52 degrees. The $\gamma$-ray energies from both the beam-like and target-like fragments were measured using the CLARA array \cite{clara}, in this experiment consisting of 23 Compton suppressed clover detectors.

\begin{figure}[h]
\begin{center}
 \epsfig{file=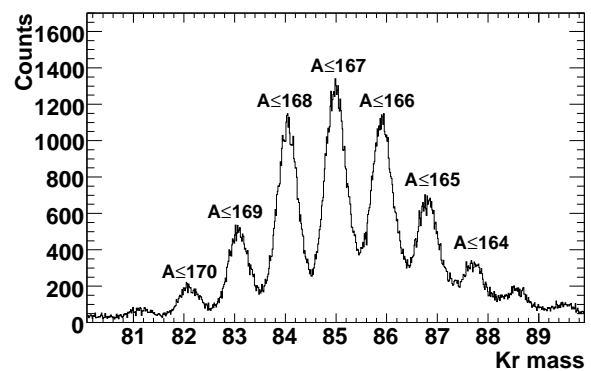, width=\columnwidth}
 \vspace{-1cm}
\end{center}
\caption{\it
Mass spectrum from PRISMA of target-like fragments gated on krypton. The masses ($A$) of the corresponding dysprosium isotopes are also shown.
}
\label{label_fig_mass}
\end{figure}

The nuclei of interest correspond to two-proton stripping ($^{168}$Dy) and two-proton stripping plus two-neutron pickup ($^{170}$Dy). Using the PRISMA information about the projectile-like fragments, the binary reaction partner can be uniquely identified. This binary reaction partner sets, due to neutron evaporation, an upper limit on the mass of the target-like fragment. The mass spectrum of the krypton isotopes is shown in fig.~\ref{label_fig_mass}, together with the corresponding dysprosium masses.

\section{Preliminary results}

\begin{figure}[h]
\begin{center}
 \epsfig{file=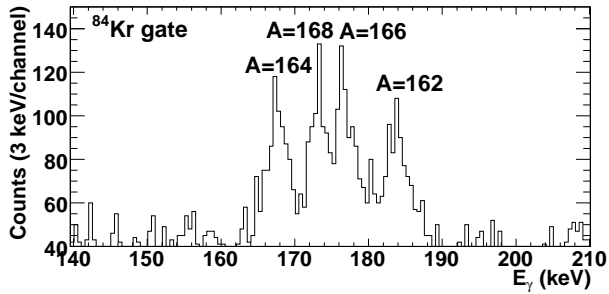, width=\columnwidth}
 \vspace{-1cm}
\end{center}
\caption{\it
Target-like $\gamma$-ray spectrum gated on $^{84}Kr$ in the energy region around the $4^{+}\to2^{+}$ transitions of the neutron rich dysprosium isotopes.
}
\label{kr84gate}
\end{figure}

The velocity vector obtained event-by-event for the beam-like fragments by PRISMA was used to Doppler correct the beam-like fragments and the target-like fragments, assuming two-body reaction kinematics. Due to the neutron evaporation, the $\gamma$-ray spectra contain lines not only from the binary dysprosium isotope, but also from lighter isotopes. See fig.~\ref{kr84gate} for an example of a target-like spectrum gated on $^{84}$Kr, the binary partner of $^{168}$Dy.


\begin{figure}[h]
\begin{center}
 \epsfig{file=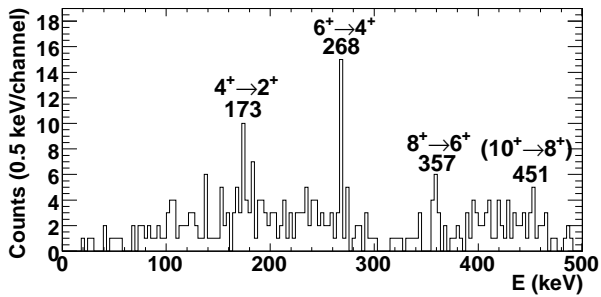, width=\columnwidth}
 \vspace{-1cm}
\end{center}
\caption{\it
Coincidence $\gamma$-ray spectrum gated on the beam-like fragments $^{82,84}$Kr and on the $\gamma$-ray energies 173~keV, 268~keV and 357~keV.
}
\label{label_fig1}
\end{figure}

To distinguish the binary channels from the neutron evaporation channels, two different methods were used. The first method was to compare the singles spectra from different gates on binary partners. For example, the $\gamma$-ray lines from $^{168}$Dy should not be visible in the spectra gated on $A>^{84}$Kr, but only in spectra gated on $A\leq^{84}$Kr. The second method was to use $\gamma\gamma$-coincidences with previously reported lines in the target-like fragments. Fig.~\ref{label_fig1} shows a $\gamma\gamma$-coincidence spectrum of $^{168}$Dy gated on the $\gamma$-ray line at 173~keV, and two previously unreported transitions at 268~keV and 357~keV.

\begin{figure}[h]
\begin{center}
 \vspace{0.5cm}
 \epsfig{file=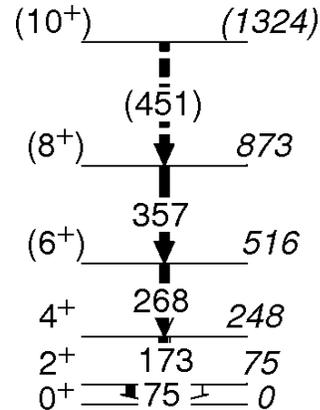, width=4cm}
\end{center}
\caption{\it
Proposed ground state rotational band of $^{168}$Dy.
}
\label{label_fig2}
\end{figure}

In the current work we have required the mutual conditions that a $\gamma$-ray line appears both in the singles spectra with the correct beam-like partners and in the $\gamma\gamma$-coincidence spectrum. The tentative $10^{+}\to8^{+}$ transition does not fulfill these requirements as it has low statistics and only appears in the $\gamma\gamma$-coincidence spectrum. See fig.~\ref{label_fig2} for a proposed partial level scheme of $^{168}$Dy. The level ordering, spins and parities have been assigned through systematics.

So far, three $\gamma$-ray lines have been identified unambiguously in $^{168}$Dy. The $4^{+}\to2^{+}$ transition at 173~keV was previously known \cite{168dy}, while the $6^{+}\to4^{+}$ transition at 268~keV and the $8^{+}\to6^{+}$ transition at 357~keV are new in this work. A tentative identification of a $10^{+}\to8^{+}$ transition at 451~keV has also been made. All the identified $\gamma$-ray lines, except the tentative $10^{+}\to8^{+}$ transition, have a lower energy than the corresponding transitions in $^{166}$Dy, implying that the increasing collectivity also occurs at higher spins. The next step in the analysis is to try to identify the corresponding lines in $^{170}$Dy, to see if this trend continues along the isotopic chain. Since no  known $\gamma$-ray lines exist in $^{170}$Dy which can be used for $\gamma\gamma$-coincidences, attempts are being made to use known $\gamma$-ray lines in the binary partner, $^{82}$Kr, as a starting point. This analysis is still ongoing.

\section{Acknowledgments}

This work was partially supported by the European Commission within the Sixth Framework Programme through I3-EURONS contract RII3-CT-2004-506065, the Swedish Research Council, EPSRC/STFC (UK) and U.S. DOE grant No. DE-FG02-91ER40609.

\end{document}